\documentclass [11pt,a4paper]{article}
\usepackage{amssymb}
\usepackage{enumerate,verbatim}
\usepackage{amsmath}
\usepackage{amsfonts,mathrsfs}
\usepackage{amsthm,wasysym}
\usepackage{epsfig,lpic,tikz}
\usepackage[applemac,utf8]{inputenc}

\usepackage[width=0.9\textwidth]{caption}

\usepackage{pdfsync}
\usepackage{xcolor,hyperref}
\usepackage[T1]{fontenc}
\usepackage{mathtools}

\hypersetup{
    colorlinks=true,
    linkcolor=blue,
    filecolor=magenta,
    urlcolor=blue,
    pdftitle={Overleaf Example},
    pdfpagemode=FullScreen,
    }

\newcounter{teoremaganso}
\newcounter{appendix}

\newcounter{coryganso}

\renewenvironment{abstract}{\small\quotation\noindent
 {\bfseries \abstractname .}}{\endquotation \par}

\newenvironment{trivlistparm}[2]{\trivlist
\item[{#1 #2}]}{\endtrivlist}

\newenvironment{prooftext}[1]{\trivlistparm{\bfseries}{#1}}{\Qed\endtrivlistparm}
\newenvironment{prova}{\trivlistparm{\bfseries}{Proof.}}{\Qed\endtrivlistparm}

\catcode`\@=11

\def\resetthefootnote{\renewcommand{\thefootnote}{\@arabic\c@footnote} }
\def\@principiremex#1{\trivlist
 \item[\hskip \labelsep{\bfseries #1\ \thetheo.}]\ignorespaces}
\def\opar@principiremex#1[#2]{\trivlist
 \item[\hskip \labelsep{\bfseries #1\ \thetheo\ (#2).}]\ignorespaces}

\newcommand{\newTHEOremrom}[2]{\newenvironment{#1}{\refstepcounter{theo}\@ifnextchar[{\opar@principiremex{#2}}
{\@principiremex{#2}}}{\qedB\endtrivlist}} \catcode`\@=12
\DeclareMathSymbol{\square}{\mathord}{AMSa}{"03}
\newcommand{\qedB}{\nopagebreak\hspace*{\fill}$\square$\par}
\newcommand{\Qed}{\nopagebreak\hspace*{\fill}{\vrule width6pt height6pt depth0pt}\par}

\theoremstyle{plain}

\newtheorem{theorem}{Theorem}[section]
\newtheorem{lemma}[theorem]{Lemma}

\newtheorem {bigtheo} [teoremaganso] {Theorem}

\newcommand{\br}{\mathbb R}

\newcommand{\bn}{\mathbb N}

\newsavebox{\savepar}

\newcommand{\refc}[1]{\mbox{$(\ref{#1})$}}

\newcommand{\teoc}[1]{Theorem~\ref{#1}}

\newcommand{\lemc}[1]{Lemma~\ref{#1}}

\newcommand{\figc}[1]{Figure~\ref{#1}}

\title{\bf Bifurcation analysis of the Microscopic Markov Chain Approach to contact-based epidemic spreading in networks}

\author{ Alex Arenas, Antonio Garijo, Sergio G\'omez and Jordi Villadelprat
\\*[.1truecm]
{\small \textsl{Departament d'Enginyeria Inform{\`a}tica i Matem{\`a}tiques,}}
\\*[-.05truecm]
{\small \textsl{Universitat Rovira i Virgili, 43007 Tarragona, Spain}}}
\begin{document}

\maketitle
\begin{abstract}
The dynamics of many epidemic compartmental models for infectious diseases that spread in a single host population present a second-order phase transition. This transition occurs as a function of the infectivity parameter, from the absence of infected individuals to an endemic state. Here, we study this transition, from the perspective of dynamical systems, for a discrete-time compartmental epidemic model known as Microscopic Markov Chain Approach, whose applicability for forecasting future scenarios of epidemic spreading has been proved very useful during the COVID-19 pandemic. We show that there is an endemic state which is stable and a global attractor and that its existence is a consequence of a transcritical bifurcation. This mathematical analysis grounds the results of the model in practical applications.
\end{abstract}

\section{Introduction and main results}\label{sec:intro}

The problem of modeling the spread of a contagious disease among individuals has been studied in deep over many years ~\cite{anderson1992infectious,hethcote2000mathematics,daley2001epidemic,pastor2015epidemic}.
The development of compartmental models, i.e., models that divide the individuals among a set of possible states, has given rise to a new collection of techniques that enable, for instance, the analysis of the onset of epidemics~\cite{pastor2001epidemic,newman2002spread,hufnagel2004forecast,wang2003epidemic,chakrabarti2008epidemic,gomez2010discrete,gomez2011nonperturbative,brockmann2013hidden,cai2014effective,cai2016solving,nowzari2016analysis}, the study of epidemics in structured networks~\cite{granell2013dynamical,granell2014competing,de2016physics,gomez2018critical,matamalas2018effective,arenas2020modeling}, or the study of the impact of a vaccination campaign~\cite{earn2000simple,pastor2002immunization,madar2004immunization,gomez2006immunization,hebert2013global,steinegger2020pulsating}. All previous works heavily rely on the mathematical approach to the study of epidemic spreading~\cite{lofgren2014mathematical} and here we follow the same spirit.

In this paper we consider a connected undirected network $\mathcal N_n$ made up of $n$ nodes, whose weights $r_{ij}\in [0,1]$ represent the contact probability between nodes $i$ and $j$. Since the network is undirected and connected, the $n\times n$ contacts matrix $R=(r_{ij})$ is symmetric and irreductible. We also assume the absence of self-loops, thus $r_{ii}=0$ for all $i$.
The non-zero entries of matrix $R$ represent the existing links in the network that are used to transmit the infection, while $r_{ij}=r_{ji}=0$ is used to indicate that nodes~$i$ and~$j$ are not connected. In the special case that all non-zero contact probabilities are one, $r_{ij}=r_{ji}=1$, matrix~$R$ becomes the adjacency matrix of the network.
Note that, for a non-connected network, we can apply our results separately to every connected component of the network.

We now define a discrete dynamical system based on the infection process on the network \cite{gomez2010discrete}, called the \emph{Microscopic Markov Chain Approach} (MMCA), that is a mathematical model for the well-known susceptible-infected-susceptible (SIS) epidemic spreading model. In the SIS model on networks, each node may be in one of two different states: susceptible (healthy) or infected. The discrete-time dynamic of the SIS makes that, at each time step, susceptible nodes may get infected (with probability $\beta\in[0,1]$) by contacts with their infected neighbours, while infected nodes may recover spontaneously (with probability $\mu\in[0,1]$). We consider that, at each time step, all nodes contact to all their neighbours, known as a {\it reactive} process. Other options are also possible, like contacting only a maximum number of neighbours, or even just one neighbour per time step; this last option is known as a {\it contact} process. From now on, we will restrict our analysis to the reactive process, which is the most common choice in the literature of the SIS model.

Following \cite{gomez2010discrete}, we also add to the SIS dynamic the possibility of one-step reinfections, which means that an infected node that has recovered, may become infected by its neighbours within the same time step. The rationale is that the recovery of a node cannot last too long if it has many infected neighbours, thus it should effectively be equivalent to a non-recovery. An example could be computer viruses and other kinds of malware: to get rid of the virus, you cannot just remove it from one computer, since the neighbours would infect it again almost immediately.

The MMCA model provides a mathematical description of the SIS spreading process based on the use of the probabilities of the nodes of being infected. Denoting $p_i^k$ the probability that node $i$ is infected at the time step $k$, its evolution is given by the MMCA equation
\begin{equation}
p_i^{k+1}= (1-q_i^k)(1-p_i^k) + (1-\mu) p_i^k + \mu (1-q_i^k) p_i^k\,,
\label{eq:pik}
\end{equation}
where $q_i^k$  is the probability that the node $i$ is not infected by any neighbor at time step $k$, whose value can be approximated as
\begin{equation}
q_i^k =\prod_{j=1}^n (1- \beta r_{ij} p_j^k)\,.
\label{eq:qik}
\end{equation}
The three summands in the right hand side of the equation in \refc{eq:pik} account for the three different ways in which a node may be infected at time $k+1$: $i)$ being susceptible at time $k$ and getting infected by its neighbours; $ii)$ being infected and not recovering; or $iii)$ being infected, recover, and becoming infected again (one-step reinfection). On the other hand, the equality in \refc{eq:qik} states that the probability of a node not being infected by any of its neighbours is equal to the product of the probabilities that each individual neighbour does not infect it. It implicitly assumes independence between the neighbours, which is good approximation in many cases, as shown in \cite{gomez2010discrete}; see \cite{matamalas2018effective} for an extension of the MMCA model that takes into account joint probabilities between pairs of connected nodes, thus significantly alleviating the independence approximation.

According to the MMCA equation in \refc{eq:pik}, the evolution of this discrete dynamical system is governed by the iteration of the map
\[
 F=(F_1,\ldots,F_n):\mathbb R^n \longrightarrow \mathbb R^n
\]
where, for $i=1,2,\ldots,n$, and setting $\mathbf{p}=(p_1,p_2,\ldots,p_n)\in\br^n,$
 \begin{equation}\label{def_F}
  F_i(\mathbf{p})\!:=  1- \big(1-(1-\mu) p_i\big) q_i(\mathbf{p})\text{ with }q_i (\mathbf{p})\!:= \prod_{j=1}^n (1- \beta r_{ij} p_j).
 \end{equation}
In other words, if $(p_1^0, \ldots, p_n^0)$ is the vector of initial conditions then $(p_1^{k+1}, \ldots. p_n^{k+1}) = F^k(p_1^0, \ldots, p_n^0)$ where $F^k=F\circ F^{k-1}$. Due to the physical nature of the problem, $F$ maps $[0,1]^n$ to $[0,1]^n$, and we restrict the study of the discrete dynamical system generated by $F$ on the compact set $\Omega=[0,1]^n$.

\begin{figure}[t]
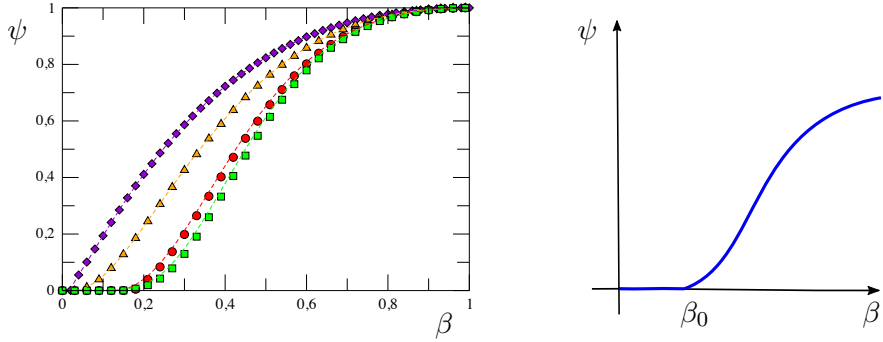

 \centering
   \begin{lpic}[l(0mm),r(0mm),t(0mm),b(5mm)]{fig01(0.75)}
    \lbl[l]{70,-2;$\beta$}
    \lbl[l]{145,0;$\beta$}
    \lbl[l]{113,0;$\beta_0$}
    \lbl[l]{95,50;$\psi$}
    \lbl[l]{-5,50;$\psi$}
   \end{lpic}
   \caption{Expected fraction of infected nodes,  $\psi\!:=\frac{1}{n}\sum_{i=1}^np_i^\infty$, as a function of the infection probability $\beta$. On the left, numerical results obtained in \cite{gomez2010discrete} by using Monte Carlo simulations (symbols) and MMCA (lines) for scale-free networks with $n=10^4$ nodes and different exponents of the degree distribution: 2.3 (purple), 2.7 (yellow), 3.3 (red), 3.5 (green). On the right, sketch showing the epidemic threshold $\beta_0$.}
    \label{psi}
\end{figure}

Numerical simulations \cite{gomez2010discrete} show that these kind of systems, governed by the map $F$ in \refc{def_F},  converge to an  asymptotic distribution
 \[
  \lim_{k \to \infty} F^k (\mathbf{p})=\mathbf{p^\infty}=(p_1^{\infty}, \ldots, p_n^{\infty})
 \]
independently on the initial condition $\mathbf{p}\in\Omega$.
Hence it seems that there exists a fixed point that is a global attractor for the discrete dynamical system under consideration. The numerical simulations also show that the location of this global attractor $\mathbf{p^\infty}$ undergoes a bifurcation at $\beta_0\!:=\frac{\mu}{\rho(R)},$ where $\rho(R)$ is the spectral radius of the matrix $R$, see \figc{psi}. Our goal in the present paper is to prove this analytically.

\begin{figure}[t]
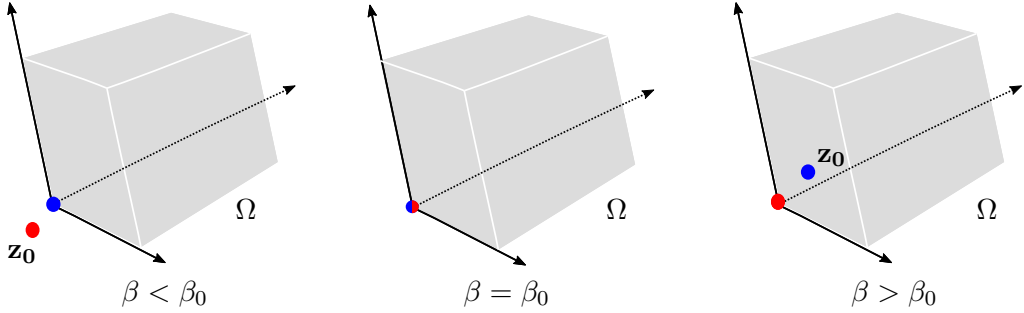

    \centering
       \begin{lpic}[l(0mm),r(0mm),t(0mm),b(8mm)]{fig02(0.75)}
           \lbl[l]{20,-5;$\beta<\beta_0$}
           \lbl[l]{80,-5;$\beta=\beta_0$}
           \lbl[l]{148,-5;$\beta>\beta_0$}
           \lbl[l]{40,10;$\Omega$}
           \lbl[l]{105,10;$\Omega$}
           \lbl[l]{170,10;$\Omega$}
           \lbl[l]{0,2;$\mathbf{z_0}$}
           \lbl[l]{142,19;$\mathbf{z_0}$}
      \end{lpic}
           \caption{\small{Sketch of the transcritical bifurcation of $F$ at $\beta_0\!:=\frac{\mu}{\rho(R)}$. For
           $\beta\approx\beta_0$ the fixed point in red is unstable and the one in blue stable.}}
    \label{figtrans}
    \end{figure}

One can easily verify that the origin ${\bf 0}=(0, \ldots, 0)$ is a fixed point of $F$ for any $\beta,\mu\in [0,1]$. We shall prove that for each $\mu\in (0,1)$ this fixed point undergoes a \emph{transcritical bifurcation} at the \emph{epidemic threshold} $\beta_0\!:=\frac{\mu}{\rho(R)},$ see \figc{figtrans}. Indeed, the origin is a stable fixed point for $\beta<\beta_0$ and, as $\beta$ tends to $\beta_0$, it collides with an unstable fixed point $\mathbf{z_0}$ coming from outside $\Omega.$ Then, for $\beta>\beta_0$, the origin is unstable while $\mathbf{z_0}$ is stable and inside $\Omega.$ This exchange of stability due to the transcritical bifurcation explains the graph in \figc{psi} because additionally we will prove that ${\bf 0}$ is a global attractor for $\beta<\beta_0$ and~$\mathbf{z_0}$ is a global attractor for $\beta>\beta_0$, i.e.,
\[
 \lim_{k \to \infty} F^k (\mathbf{p})=\left\{
  \begin{array}{ll}
   {\bf 0} & \text{ if $\beta<\beta_0$},\\
   \mathbf{z_0} & \text{ if $\beta>\beta_0$,}
  \end{array}
 \right.
 \text{for all ${\bf p}\in\Omega\setminus\{{\bf 0}\}.$}
\]
More concretely, our main results are summarized in the following statement, where $\|\;\|_2$ stands for the Euclidean norm in $\br^n.$

\begin{bigtheo}\label{teoN}
Let us consider a connected undirected network $\mathcal N_n$  with associated matrix $R$ and parameters $\beta, \mu \in (0,1)$. Then the following holds:
\begin{enumerate}[$(a)$]

\item The origin $\mathbf{0}$ is a fixed point of $F$ for all parameter value and, for each $\mu$, it undergoes
         a transcritical bifurcation as the $\beta$ varies through the bifurcation value $\beta_0\!:=\frac{\mu}{\rho(R)}.$

\item If $\beta<\beta_0$ then $\mathbf{0}$ is a stable hyperbolic fixed point of $F$ and $\lim_{k \to \infty} F^k(\mathbf{x})= \mathbf{0}$ for all
         $\mathbf{x}\in [0,1]^n$.

\item If $\beta>\beta_0$ then there exists a fixed point $\mathbf{z}_0$ of $F$ in the interior of $[0,1]^n$ that is stable
         and verifying $\lim_{k \to \infty} F^k(\mathbf{x})= \mathbf{z}_0$ for all
         $\mathbf{x} \in [0,1]^n\setminus\{\mathbf{0}\}$. Moreover the map $\beta\mapsto\|\mathbf{z}_0\|_2$
         is monotonous increasing.


\end{enumerate}
\end{bigtheo}

In the proof of \teoc{teoN} we combine local and global techniques. The most difficult part is of course to prove the global attraction of a local attracting fixed point. For the parameter values in statement $(b)$ the map $F$ is contracting on $[0,1]^n$ and the result follows by applying the Contraction Mapping Theorem. The same approach is no longer valid in order to show $(c)$ because for those parameter values the map $F$ has two different fixed points on $[0,1]^n$. We use in this case that the well-known facts about fixed points of positive, monotone and convex functions on the real line extend to similar maps on an \emph{ordered Banach space} (i.e., a Banach space with a partial order induced by a positive cone).

\section{ Proof of the main results}\label{proves}

The following result is well known (see for instance \cite[pp. 154]{ortega1990numerical}) but since we were not able to find a proof we include it here for completeness.
In the statement $(DG)_x$ stands for the differential matrix of $G$ at the point $x$. We also consider the
vector $p$-norm $\|x\|_p$ in $\br^n$, $1\leqslant p\leqslant\infty$, and its induced matrix norm $\|A\|_p=\sup_{x\neq 0}\frac{\|Ax\|_p}{\|x\|_p}$ in $M_{n\times n}.$

\begin{lemma}\label{cota}
Let $D$ be a convex subset of $\br^n$ and consider a $\mathscr C^1$ mapping $G\!:\!D\to\br^n$ such that $\|DG_x\|_p\leqslant\kappa$ for all $x\in D.$ Then $\|G(x)-G(y)\|_p\leqslant\kappa\|x-y\|_p$ for all $x,y\in D.$
\end{lemma}

\begin{prova}
Given $x,y\in D$, let us set $g(t)\!:=G\big(tx+(1-t)y\big),$ which is a well defined $\mathscr C^1$ function from the interval $[0,1]$ to~$\br^n,$ so that
\[
 G(x)-G(y)=g(1)-g(0)=\int_0^1g'(t)dt=\int_0^1\big(DG\big)_{tx+(1-t)y}(x-y)dt.
\]
Consequently
\[
 \|G(x)-G(y)\|_p\leqslant \int_0^1\|\big(DG\big)_{tx+(1-t)y}(x-y)\|_pdt\leqslant\kappa\|x-y\|_p,
\]
where in the second inequality we use that $\|Ax\|_p\leqslant \|A\|_p\|x\|_p$. This proves the result.
\end{prova}

In the next statement, and in what follows, we say that $A=(a_{ij})$ is a \emph{nonnegative} (respectively, \emph{positive}) matrix if $a_{ij}\geqslant 0$ (respectively, $a_{ij}>0)$ for all $i,j$.  We also
define entrywise inequalities for two matrices $A=(a_{ij})$ and $B=(b_{ij})$ with the same size as
\begin{align}\label{newleq}
 &A\preceq B \Leftrightarrow \text{$a_{ij}\leqslant b_{ij}$ for all $i,j$}\\
 \intertext{and}\notag
 &A\prec B \Leftrightarrow \text{$a_{ij}< b_{ij}$ for all $i,j$.}
\end{align}
The reverse relations $A\succeq B$ and $A\succ B$ are defined similarly \cite{gentle2007matrix}.

\begin{lemma}\label{spec_norm}
The following holds:
\begin{enumerate}[$(a)$]
 \item If $A$ and $B$ are nonnegative square matrices with $A\preceq B$ then $\rho(A)\leqslant \rho(B)$ and
          $\|A\|_2\leqslant \|B\|_2.$
 \item If $A$ is a nonnegative square matrix then $\rho(Id+A)=1+\rho(A).$
 \item If $A$ is a symmetric matrix then $\|A\|_2=\rho(A).$
 \item If $A$ is a nonnegative square matrix then $\rho(A)$ is an eigenvalue of $A$ and there is a
          nonnegative vector $\mathbf{u}\neq\mathbf{0}$ such that $A\mathbf{u}=\rho(A)\mathbf{u}.$ Moreover
          the algebraic multiplicity of the eigenvalue $\rho(A)$ is 1 in case that $A$ is an irreductible matrix.
 \end{enumerate}
\end{lemma}

\begin{prova}
All the assertions are well-known and we refer the reader to \cite{horn1985matrix} for the proof. More concretely, for the first and second inequality in $(a)$ see Corollary 8.1.19 and 5.6.P41, respectively. The assertion in $(b)$  is proved in Lemma 8.4.2. On the other hand $\|A\|_2=\sqrt{\rho(A^tA)},$ see page 346, so that $(c)$ follows using that~$A$ is symmetric by assumption. The proof of the first assertion in~$(d)$ can be found in Theorem 8.3.1, whereas the second one follows by the Perron-Frobenius Theorem (see Theorem 8.4.4).
\end{prova}

\begin{figure}[tb!]
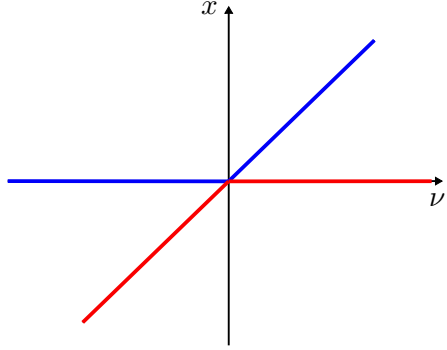

 \centering
   \begin{lpic}[l(0mm),r(0mm),t(0mm),b(5mm)]{fig03(0.75)}
    \lbl[l]{74,26;$\nu$}
    \lbl[l]{34,60;$x$}
   \end{lpic}
   \caption{Transcritical bifurcation in the model $x\mapsto (1+\nu)x-x^2$, where the stable fixed point is depicted in blue and the unstable one in red.}
    \label{fig:transcritical}
\end{figure}

The following result characterizes the so-called \emph{transcritical bifurcation}. The toy model for this kind of 1-parameter bifurcation is the iteration of the map $x\mapsto (1+\nu)x-x^2$, which has two fixed points, one at $x=0$ for all $\nu$ and the other at $x=\nu,$ see \figc{fig:transcritical}. For $\nu<0$ the fixed point $x=0$ is stable, whereas $x=\nu$ is unstable. As $\nu$ increases, the unstable fixed point approaches the origin and coalesces with it when $\nu=0$. Finally, when $\nu>0$ the origin becomes unstable and $x=\nu$ is now stable. In other words there has been an exchange of stabilities between the two fixed points. The next result provides sufficient conditions for the occurrence of this local bifurcation when the phase space is $n$-dimensional. Its counterpart for flows was originally proved by J. Sotomayor in \cite{sotomayor1973generic}, see also \cite[p. 150]{guckenheimer1983nonlinear} and \cite[p. 338]{perko2013differential}. With regard to the version for iteration of maps the reader is referred to \cite[chapter VII]{robinson1998dynamical}, where it is given the proof of a similar result for the saddle-node bifurcation that can be easily adapted.

\begin{theorem}\label{trans}
Let $f\!:\!\br^n\times\br\rightarrow\br^n$ be a $\mathscr C^2$ map verifying the following:
\begin{enumerate}[$(a)$]

 \item $\mathbf{x}_0$ is a fixed point for all $\nu,$ i.e., $f(\mathbf{x}_0;\nu)=\mathbf{x}_0$ for all $\nu.$

 \item The Jacobian matrix of $f(\,\cdot\,;\nu_0)$ evaluated at $\mathbf{x}=\mathbf{x}_0$, that is
          $D_{\mathbf{x}}f(\mathbf{x}_0;\nu_0)$, has a simple eigenvalue $\lambda=1$
          and all the other eigenvalues have
          modulus strictly smaller than one.

\item  The derivatives
 \begin{align*}
&\mathbf{w}\left[D_{\mathbf{x}\mathbf{x}}f(\mathbf{x}_0;\nu_0)(\mathbf{v},\mathbf{v})\right]
     =\sum_{i,j,k=1}^nw_kv_iv_j\frac{\partial^2f_k(\mathbf{x}_0;\nu_0)}{\partial x_i\partial x_j}
     \\
\intertext{and}
&\mathbf{w}\left[D_{\mathbf{x}\nu}f(\mathbf{x}_0;\nu_0)\mathbf{v}\right]=\sum_{i,k=1}^nw_kv_i\frac{\partial^2f_k(\mathbf{x}_0;\nu_0)}{\partial x_i\partial\nu}
\end{align*}
are different from zero, where $\mathbf{v}$ and $\mathbf{w}$ are respectively the
right $($column$)$ and left $($row$)$ eigenvectors for $\lambda=1$ of $D_{\mathbf{x}}f(\mathbf{x}_0;\nu_0)$.

\end{enumerate}
Then the discrete dynamical system that yields the iteration of the map $\mathbf{x}\mapsto f(\mathbf{x};\nu)$
undergoes a transcritical bifurcation at the fixed point $\mathbf{x}_0$ as $\nu$ varies through the bifurcation value $\nu=\nu_0.$
\end{theorem}

\begin{prooftext}{Proof of \teoc{teoN}.}
Our first task will be to compute the first and second order partial derivatives of the map $F=(F_1,\ldots,F_n):\br^n\rightarrow\br^n.$ Recall that, for each $i=1,2,\ldots,n,$
\begin{equation}\label{teoNeq0}
  F_i(\mathbf{x}) =  1+ \big((1-\mu) x_i-1\big) q_i(\mathbf{x})\text{, where }
  q_i (\mathbf{x})=\prod_{j=1}^n (1- \beta r_{ij} x_j).
 \end{equation}
Let us first note that, for $j =1, \ldots, n,$
\[
\frac{ \partial q_i (\mathbf{x})}{ \partial x_j}
=- \beta r_{ij} \prod_{\substack{k=1 \\ k\neq j}}^n (1- \beta r_{ik} x_k),
\]
which vanishes in case that $i=j$ since $r_{ii}=0$ by assumption. Hence
\begin{equation}\label{teoNeq1}
\frac{\partial F_i (\mathbf{x})}{\partial x_i}  =  (1-\mu) q_i (\mathbf{x})
\text{ and, for $j \neq i,$ }
\frac{ \partial F_i (\mathbf{x})}{\partial x_j} =\beta r_{ij}\big(1-(1-\mu) x_i\big)\prod_{\substack{k=1 \\ k\neq j}}^n (1- \beta r_{ik} x_k).
\end{equation}
Thus the Jacobian matrix of $F$ at the origin writes as
\begin{equation}\label{teoNeq8}
D_{\mathbf{x}}F(\mathbf{0})=
\left(
\begin{array} {cccc}
1-\mu & \beta r_{12} & \ldots & \beta r_{1n} \\
\beta r_{21} & 1-\mu &  \ldots & \beta r_{2n} \\
      &  & & \\
\vdots & \vdots & \ddots & \vdots \\
  &  & & \\
 \beta r_{n1} & \ldots & \beta r_{n \, n-1} & 1-\mu
\end{array}
\right)
=(1-\mu)Id+\beta R.
\end{equation}
Some easy computations show that if $i,$ $j$ and $k$ are pairwise distinct then
\begin{equation}\label{teoNeq13}
\begin{array}{lll}
 \displaystyle\frac{\partial^2F_i(\mathbf{x})}{\partial x_j\partial x_k}=-\beta^2 r_{ij}r_{ik}(1-(1-\mu)x_i)\prod_{\substack{\ell=1 \\ \ell\neq j,k}}^n (1- \beta r_{i\ell} x_\ell)
 & &  \displaystyle\frac{\partial^2F_i(\mathbf{x})}{\partial x_i^2}=0
 \\
 \displaystyle\frac{\partial^2F_i(\mathbf{x})}{\partial x_i\partial x_j}=-\beta r_{ij}(1-\mu)\prod_{\substack{\ell=1 \\ \ell\neq j}}^n (1- \beta r_{i\ell} x_\ell)  &\quad&
 \displaystyle\frac{\partial^2F_i(\mathbf{x})}{\partial x_j^2}=0 \\
\end{array}
\end{equation}
In particular,
\begin{equation}\label{teoNeq7}
 \frac{\partial^2F_i(\mathbf{0})}{\partial x_j\partial x_k}=-\beta^2 r_{ij}r_{ik}\leqslant 0,
 \text{ }\frac{\partial^2F_i(\mathbf{0})}{\partial x_j^2}=0\text{ and }\frac{\partial^2F_i(\mathbf{0})}{\partial x_i\partial x_j}=-\beta r_{ij}(1-\mu)\leqslant 0.
\end{equation}

That being established, we begin with the proof of assertion $(a)$, that will follow by applying \teoc{trans}.
To this end we fix $\mu$ and to stress the dependence on $\beta$ we introduce the notation $F(\mathbf{x};\beta).$ In doing so we observe that $F(\mathbf{0};\beta)=\mathbf{0}$ for all $\beta$ and, from \refc{teoNeq8},
\begin{equation}\label{teoNeq2}
 \rho\big(D_{\mathbf{x}}F(\mathbf{0;\beta})\big)=1-\mu+\beta\rho(R).
\end{equation}
Here we also apply $(b)$ in \lemc{spec_norm} taking $1-\mu>0$ and $\beta>0$ into account. It is also clear that $\mathbf{v}$ is an eigenvector of $R$ with eigenvalue~$\lambda$ if, and only if, $\mathbf{v}$ is an eigenvector of $D_{\mathbf{x}}F(\mathbf{0};\beta)$ with eigenvalue $1-\mu+\beta\lambda.$ Note in addition that the multiplicities of the respective eigenvalues are the same. By $(d)$ in \lemc{spec_norm}, since $R$ is an irreducible and nonnegative matrix, $\lambda=\rho(R)$ is a simple eigenvalue of $R$ and there is a nonnegative vector $\mathbf{u}\neq 0$ such that
\begin{equation}\label{teoNeq11}
 R\mathbf{u}=\rho(R)\mathbf{u}.
\end{equation}
Accordingly if we set $\beta_0\!:=\frac{\mu}{\rho(R)}$, from \refc{teoNeq2}, the Jacobian matrix $D_{\mathbf{x}}F(\mathbf{0};\beta_0)$ has a simple eigenvalue $\lambda=1$ and all the other eigenvalues have modulus strictly smaller than one. In particular,
\begin{equation}\label{teoNeq10}
D_{\mathbf{x}}F(\mathbf{0};\beta_0)\mathbf{u}=1\mathbf{u}.
\end{equation}
So far we have proved that the assumptions $(a)$ and $(b)$ in \teoc{trans} hold. In order to show that~$(c)$ is also true we note that $D_{\mathbf{x}}F(\mathbf{0};\beta_0)$ is a symmetric matrix, see \refc{teoNeq8}, so that its right and left eigenvectors of $\lambda=1$ are, respectively, $\mathbf{u}$ and $\mathbf{u}^t.$ Taking this into account we claim that
\begin{equation}\label{teoNeq9}
 \mathbf{u}^t\left[D_{\mathbf{x}\mathbf{x}}F(\mathbf{0};\beta_0)(\mathbf{u},\mathbf{u})\right]
     =\sum_{i,j,k=1}^nu_ku_iu_j\frac{\partial^2F_k(\mathbf{0};\beta_0)}{\partial x_i\partial x_j}<0.
\end{equation}
To this end we shall use that $\mathbf{u}$ is a nonnegative vector. Therefore $\mathbf{u}=(u_1,\ldots,u_n)$ with $u_i\geqslant 0$ for all $i=1,2,\ldots,n$ and there exists some $\ell$ such that $u_\ell>0.$ On account of \refc{teoNeq7}, the claim will follow once we prove that the sum of the $n$ terms in~\refc{teoNeq9} with $k=i=\ell$ is exactly $-\mu(1-\mu)u_\ell^3,$ which is negative.
Indeed, from~\refc{teoNeq8} and~\refc{teoNeq10}, we get
\[
 \beta_0\sum_{j=1}^nr_{\ell j}u_j=\mu u_\ell>0,
\]
and the combination of this with the third equality in \refc{teoNeq7} yields
\[
 u_\ell^2\sum_{j=1}^nu_j\frac{\partial^2F_\ell(\mathbf{0};\beta_0)}{\partial x_\ell\partial x_j}
 =-\beta_0(1-\mu)u_\ell^2\sum_{j=1}^nr_{\ell j}u_j=-\mu(1-\mu) u_\ell^3<0.
\]
This proves the inequality in \refc{teoNeq9}, as desired. On the other hand, from \refc{teoNeq8} and \refc{teoNeq11},
\[
 \mathbf{u}^t\left[D_{\mathbf{x}\beta}F(\mathbf{0};\beta_0)\mathbf{u}\right]=\frac{\mu}{\beta_0}\|\mathbf{u}\|^2\neq 0,
\]
where we also use that $\beta_0=\frac{\mu}{\rho(R)}$ by definition. This shows that the last assumption in \teoc{trans}  is also satisfied and so we can conclude that the fixed point at the origin undergoes a transcritical bifurcation as the $\beta$ varies through the bifurcation value $\beta_0.$

From now on, for simplicity in the exposition, we shall omit the dependence of $F$ on the parameters. That being said,
let us turn now to the proof of the assertions in $(b)$. With this aim in view we first note that, from \refc{teoNeq1},
\[
0\leqslant\frac{ \partial F_i (\mathbf{x})}{\partial x_j}\leqslant \beta r_{ij}
\text{, for $j\neq i$, and }
0<\frac{\partial F_i (\mathbf{x})}{\partial x_i}\leqslant 1-\mu.
\]
Here we also use that $0< 1-\beta r_{ij}x_j\leqslant 1$ due to $r_{ij},x_j\in [0,1]$. Thus, for all $\mathbf{x}\in [0,1]^n$, the Jacobian matrix $DF({\mathbf{x}})$ is a nonnegative and verifies $DF({\mathbf{x}})\preceq (1-\mu)Id+\beta R,$ recall \refc{newleq}. Hence, by applying \lemc{spec_norm},
\begin{equation*}
       \|DF({\mathbf{x}})\|_2\leqslant\| (1-\mu)Id+\beta R\|_2=\rho\big((1-\mu)Id+\beta R\big)= 1-\mu+\beta\rho(R)
       \text{ for all $\mathbf{x}\in [0,1]^n,$}
\end{equation*}
where we use that $R$ is symmetric.
It is clear then that the condition $\beta<\beta_0\!:=\frac{\mu}{\rho(R)}$ implies $\|DF({\mathbf{x}})\|_2\leqslant\kappa$ for all $\mathbf{x}\in [0,1]^n$ with $\kappa\in [0,1).$ Thus, by applying \lemc{cota} with $p=2$, $F$ is a contraction on $[0,1]^n$. Since one can easily verify that $F\big([0,1]^n\big)\subset [0,1]^n$ and $F(\mathbf{0})=\mathbf{0}$, the application of the Contraction Mapping Theorem (see for instance \cite[Theorem 2.5]{robinson1998dynamical}) shows that $\lim_{k \to \infty} F^k(\mathbf{x})= \mathbf{0}$ for all $\mathbf{x}\in [0,1]^n$. The fact that $\mathbf{0}$ is hyperbolic follows from~\refc{teoNeq2} because $\rho\big(DF({\mathbf{0}})\big)= 1-\mu+\beta\rho(R)<1$ provided that $\beta<\beta_0.$
On account of this, and by applying the Stable Manifold Theorem (see \cite[Theorem 10.1]{robinson1998dynamical} for instance), we conclude that $\mathbf{0}$ is stable.

We proceed next with the proof of $(c)$. So let us assume that $\beta>\beta_0$. In this case from~\refc{teoNeq2} it turns out that $r\!:=\rho\big(DF(\mathbf{0})\big)>1.$ Thus, since it is a nonnegative matrix, by applying $(d)$ in \lemc{spec_norm} there exists a nonnegative vector $\mathbf{v}=(v_1,\ldots,v_n)\neq\mathbf{0}$ such that
\begin{equation}\label{teoNeq3}
 DF(\mathbf{0})\mathbf{v}=r\mathbf{v}\text{ with $r>1.$}
\end{equation}
We claim that
\[
 F(\varepsilon\mathbf{v})\succeq\varepsilon\mathbf{v}\text{ for $\varepsilon>0$ small enough.}
\]
In order to prove this we use \refc{teoNeq3} and that $F(\mathbf{z})=F(0)+(DF)_{\mathbf{0}}\mathbf{z}+\mathrm{o}(\|\mathbf{z}\|),$ to get
\[
 F(\varepsilon\mathbf{v})-\varepsilon\mathbf{v}=\varepsilon r\mathbf{v}-\varepsilon\mathbf{v}+\rm{o}(\|\varepsilon \mathbf{v}\|)=\varepsilon\|\mathbf{v}\|\left(\frac{r-1}{\|\mathbf{v}\|}\mathbf{v}+\frac{\rm{o}(\|\varepsilon \mathbf{v}\|)}{\|\varepsilon \mathbf{v}\|)}\right).
\]
Since $r>1$ this shows that the $i$-th component of $F(\varepsilon\mathbf{v})-\varepsilon\mathbf{v}$ is strictly positive for $\varepsilon>0$ small enough provided that $v_i>0.$ In case that $v_i=0$ we will show that the $i$-th component of $F(\varepsilon\mathbf{v})$ is also zero. Indeed, from \refc{teoNeq3} we obtain
\[
 (1-\mu)v_k+\beta\sum_{j=1}^nr_{kj}v_j=rv_k\text{ for $k=1,2,\ldots,n,$}
\]
which particularised to $k=i$ yields $\sum_{j=1}^nr_{ij}v_j=0$. Since all the summands are nonnegative this implies that $r_{ij}v_j=0$ for all $j=1,2,\ldots,n$. Thus, from \refc{teoNeq0}, $F_i(\varepsilon\mathbf{v})=0,$ as desired. This proves the validity of the claim. We observe on the other hand that the image of $\mathbf{1}\!:=(1,\ldots,1)$ by $F$ is in the interior of $[0,1]^n.$ In short, shrinking $\varepsilon>0$ if necessary and setting $\mathbf{x}_\varepsilon\!:=\varepsilon\mathbf{v}$, we have that
\begin{equation}\label{teoNeq4}
 \mathbf{x}_\varepsilon\preceq F(\mathbf{x}_\varepsilon)\prec F(\mathbf{1})\prec\mathbf{1}.
\end{equation}
Our next goal will be to prove that, for any $\mathbf{x},\mathbf{y}\in [0,1]^n,$
\begin{equation}\label{teoNeq5}
 \mathbf{x}\preceq\mathbf{y}\Rightarrow F(\mathbf{x})\preceq F(\mathbf{y}).
\end{equation}
Indeed, to see this let us set $g(t)\!:=F\big(t\mathbf{y}+(1-t)\mathbf{x}\big),$ which is a well defined $\mathscr C^1$ function from $[0,1]$ to~$\br^n$, so  we can write
\[
 F(\mathbf{y})-F(\mathbf{x})=g(1)-g(0)=\int_0^1g'(t)dt=\int_0^1\big(DF\big)_{t\mathbf{y}+(1-t)\mathbf{x}}(\mathbf{y}-\mathbf{x})dt.
\]
The integrand is a nonnegative vector because we have already proved that the Jacobian of $F$ at any $\mathbf{x}\in [0,1]^n$ is a nonnegative matrix and, on the other hand, $\mathbf{x}\preceq\mathbf{y}$ by assumption. This fact implies $F(\mathbf{x})\preceq F(\mathbf{y})$ and proves the validity of the assertion in \refc{teoNeq5}.

Given any two points $\mathbf{a},\mathbf{b}\in [0,1]^n$ with $\mathbf{a}\preceq\mathbf{b}$  we define the hypercube
\[
 \Omega(\mathbf{a},\mathbf{b})\!:=\{\mathbf{z}\in [0,1]^n:\mathbf{a}\preceq \mathbf{z}\preceq\mathbf{b}\}.
\]
Then, since $\mathbf{x}_\varepsilon\prec\mathbf{z}\prec\mathbf{1}$ implies $\mathbf{x}_\varepsilon\preceq F(\mathbf{x}_\varepsilon)\preceq F(\mathbf{z})\preceq F(\mathbf{1})\preceq\mathbf{1}$ due to \refc{teoNeq4} and \refc{teoNeq5}, we obtain that
\begin{equation}\label{teoNeq17}
 F^k\Big(\Omega(\mathbf{x}_\varepsilon,\mathbf{1})\Big)\subset\Omega\left(F^k(\mathbf{x}_\varepsilon),F^k(\mathbf{1})\right)\text{ for all $k\in\bn$}.
\end{equation}
The sequence $\{F^k(\mathbf{x}_\varepsilon)\}_{k\in\bn}$ converges to a fixed point $\mathbf{z}_0$ of $F$ inside $[0,1]^n\setminus\{\mathbf{0}\}$ because each one of the entries is a monotonous increasing sequence of real numbers smaller than $1,$ again due to~\refc{teoNeq4} and~\refc{teoNeq5}. Similarly $\{F^k(\mathbf{1})\}_{k\in\bn}$ converges to a fixed point $\mathbf{z}_1$ of $F$ inside $[0,1]^n\setminus\{\mathbf{0}\}$ because each one of the entries is a monotonous decreasing sequence of real numbers greater than $0.$
Consequently
\[
\bigcap_{k\geqslant 1}F^k\Big(\Omega(\mathbf{x}_\varepsilon,\mathbf{1})\Big)\subset\Omega\left(\mathbf{z}_0,\mathbf{z}_1\right),
\]
where $\mathbf{z}_0$ and $\mathbf{z}_1$ are fixed points of $F$ verifying
$\mathbf{0}\prec\mathbf{z}_0\preceq\mathbf{z}_1\prec\mathbf{1}$. Furthermore, since $\mathbf{x}_\varepsilon$ tends to $\mathbf{0}$ as $\varepsilon\to 0$, this shows that
\[
\bigcap_{k\geqslant 1}F^k\Big((0,1]^n\Big)\subset\Omega\left(\mathbf{z}_0,\mathbf{z}_1\right).
\]
At this point we claim that if $\mathbf{x}\in [0,1]^n\setminus\{\mathbf{0}\}$ with $\prod_{i=1}^nx_i=0$ then $F^{n-1}(\mathbf{x})\in (0,1]^n.$ Clearly, on account of the above inclusion, once we prove this we will get that
\begin{equation}\label{teoNeq14}
\bigcap_{k\geqslant 1}F^k\Big([0,1]^n\setminus\{\mathbf{0}\}\Big)\subset\Omega\left(\mathbf{z}_0,\mathbf{z}_1\right).
\end{equation}
For the sake of simplicity in the exposition, in order to prove the claim we assume, for instance, that $x_1\neq 0$. To this aim let us also note, see \refc{teoNeq0}, that the $i$-th component~$F_i(\mathbf{x})$ is equal to zero if, and only if,
\[
 \big(1-(1-\mu) x_i\big) \prod_{j=1}^n (1- \beta r_{ij} x_j)=1,
\]
which in turn occurs if, and only if, $x_i=0$ and $r_{ij}x_j=0$ for all $j=1,2,\ldots,n.$ Consequently $F_1(\mathbf{x})\neq 0.$ Observe moreover that there exists at least one $j\in\{2,\ldots,n\}$ such that $r_{1j}>0$ because the node labeled by~1 must be linked with at least another node. Thus $r_{1j}x_1>0$, so that, in addition to $F_1(\mathbf{x})\neq 0,$ we can also assert that $F_{j}(\mathbf{x})\neq 0.$ Repeating this argument we obtain that $F^2(\mathbf{x})$ has at least three components different from zero, $F^3(\mathbf{x})$ has at least four components different from zero, and so on. Hence $F^{n-1}(\mathbf{x})\in (0,1]^n$. This proves the claim and, accordingly, the validity of \refc{teoNeq14}.

It is clear at this point that if we show that $\mathbf{z}_0=\mathbf{z}_1$ then, taking \refc{teoNeq14} into account,
\begin{equation}\label{teoNeq16}
\lim_{k \to \infty} F^k(\mathbf{x})= \mathbf{z}_0\text{ for all
         $\mathbf{x} \in [0,1]^n\setminus\{\mathbf{0}\}$.}
\end{equation}
To prove this we shall appeal to the results of H. Amann \cite{amann1976fixed} with regard to the fixed points in ordered Banach spaces. More concretely, the fact that $\mathbf{z}_0=\mathbf{z}_1$ follows by applying \cite[Theorem 24.3]{amann1976fixed}, which asserts if $E$ is an ordered Banach space whose positive cone $P$ has nonempty interior, $D$ is a convex subset of $E$ and $f\!:\!D\to E$ is a strongly increasing and strongly order concave map with a fixed point $x_0\in D$, then $f$ has at most one fixed point $\bar x$ with $\bar x>x_0.$ Thus our task is to show that if we take $E=\br^n$ with the usual norm, $P=\{\mathbf{x}\in\br^n:x_i\geqslant 0\text{ for all $i=1,2,\ldots,n$}\}$ and $D=[0,1)^n$ then $\left.F\right|_D$ is a strongly increasing and strongly order concave map. For readers convenience we explain succinctly the involved notions to check that the hypothesis in \cite[Theorem 24.3]{amann1976fixed} are fulfilled. The \emph{ordering} induced by a cone $P$ in $E$ is defined as
\[
 x\leqq y\Leftrightarrow y-x\in P.
\]
As usual, $x<y$ means $x\leqq y$ but $x\neq y.$ If the interior of the positive cone $P$ is nonempty, i.e., $\mathring P\neq\emptyset,$ then a map $f$ is said to be \emph{strongly increasing} (see \cite[p. 641]{amann1976fixed}) in case that
\[
 x<y\Rightarrow f(y)-f(x)\in\mathring P.
\]
In our setting, the fact that $\left.F\right|_D$ is a strongly increasing follows by applying \cite[Theorem 7.2]{amann1976fixed} because for each $\mathbf{x}\in D=[0,1)^n$ we have
\[
 (DF)_{\mathbf{x}}u\in\mathring P \text{ for all $u\in\mathring P.$}
\]
Indeed, this is an easy consequence of the following three observations:
\begin{itemize}
\item $u\in\mathring P$ if, and only if, $u_i>0$ for all $i=1,2,\ldots,n$.
\item $\frac{ \partial F_i (\mathbf{x})}{\partial x_j}\geqslant 0$ for all $\mathbf{x}\in [0,1)^n$ and, moreover, $\frac{ \partial F_i (\mathbf{x})}{\partial x_j}=0$ if, and only if, $r_{ij}=0.$
\item Every row in the matrix $R=(r_{ij})$ has at least one strictly positive entry.
\end{itemize}
Let $D$ be a nonempty convex subset of $E.$ A map $f\!:\!D\to E$ is said to be \emph{strongly order convex} if
\[
f(x)+\tau\big(f(y)-f(x)\big)- f\big(x+\tau(y-x)\big)\in\mathring P
\]
for every $\tau\in (0,1)$ and every pair of distinct comparable points $x,y\in D,$ see \cite[p. 690]{amann1976fixed}. The map is called \emph{strongly order concave} if $-f$ is strongly order convex. We shall show that $G\!:=-F|_D$ is a strongly order convex map by applying \cite[Theorem 23.3]{amann1976fixed}, which characterizes these maps in terms of a condition on the second order derivative. In our case this condition is verified if, for each $\mathbf{x}\in D=[0,1)^n$ and $i=1,2,\ldots,n,$
\begin{equation}\label{teoNeq15}
 u^tH(G_i)_{\mathbf{x}}u>0\text{ for all $u\in\mathring P,$}
\end{equation}
where $H(G_i)_{\mathbf{x}}$ is the Hessian matrix of $G_i\!:\!D\to\br$ at $\mathbf{x}$. In its regard, from~\refc{teoNeq13}, we get that $\frac{\partial^2G_i(\mathbf{x})}{\partial x_j\partial x_k}\geqslant 0$ for all $\mathbf{x}\in D=[0,1)^n$. Moreover, in case that $i\neq j$, we have $\frac{\partial^2G_i(\mathbf{x})}{\partial x_i\partial x_j}=0$ if, and only if $r_{ij}=0.$ On account of this the validity of \refc{teoNeq15} follows noting that for each $i$ there exists at least one $j\neq i$ with $r_{ij}>0$ and that if $u\in\mathring P$ then $u_i>0$ for all $i=1,2,\ldots,n$. Hence $F|_D$ is indeed a strongly order concave map.
We are now in position to apply \cite[Theorem 24.3]{amann1976fixed} to the restriction of $F$ to $D=[0,1)^n$. In doing so, due to $F(\mathbf{0})=\mathbf{0}$ and, recall \refc{teoNeq14}, $\mathbf{0}<\mathbf{z}_0$, we obtain that $\mathbf{z}_0=\mathbf{z}_1$ and, consequently, \refc{teoNeq16} follows.

We remark that, from \refc{teoNeq16}, $\mathbf{z}_0$ is the unique fixed point of $F$ on $[0,1]^n\setminus\{\mathbf{0}\}$. The fact that it is stable follows noting that, on account of \refc{teoNeq17}, $\mathbf{z}_0$ is inside a sequence of invariant hypercubes that shrink to it. It only remains to be proved that the map $\beta\mapsto\|\mathbf{z}_0\|_2$ is increasing. To this end, for each fixed $\mu$, we denote by $\mathbf{z}_0(\beta)$ the unique fixed point of $F(\,\cdot\,;\beta)$ on $[0,1]^n\setminus\{\mathbf{0}\}$. We also observe that
\[
 \frac{\partial F_i(\mathbf{x};\beta)}{\partial\beta}=(1-(1-\mu)x_i)q_i(\mathbf{x})\sum_{j=1}^n\frac{r_{ij}x_j}{1-\beta r_{ij}x_j}\geqslant 0\text{ for all $\mathbf{x}\in [0,1]^n.$}
\]
Given any $\mathbf{x}\in [0,1]^n,$ this implies that if $\beta_1\leqslant\beta_2$ then $F(\mathbf{x};\beta_1)\preceq F(\mathbf{x};\beta_2).$ Indeed, this is so because, setting $h(t)\!:=F\big(\mathbf{x};t\beta_2+(1-t)\beta_1\big)$,
\[
 F(\mathbf{x};\beta_2)- F(\mathbf{x};\beta_1)=h(1)-h(0)=\int_0^1h'(t)dt
 =(\beta_2-\beta_1)\int_0^1\left.\nabla_\beta F(\mathbf{x};\beta)\right|_{\beta=t\beta_2+(1-t)\beta_1}dt,
\]
with the integrand being a nonnegative vector due to $\partial_\beta F_i(\mathbf{x};\beta)\geqslant 0$ for all $i=1,2,\ldots,n.$ In particular this shows that if $\beta_1\leqslant\beta_2$ then
\[
  \mathbf{z}_0(\beta_1)=F\big(\mathbf{z}_0(\beta_1);\beta_1\big)\preceq F\big(\mathbf{z}_0(\beta_1);\beta_2\big),
\]
which, on account of \refc{teoNeq5}, implies that the bounded sequence $\{F^k\big(\mathbf{z}_0(\beta_1);\beta_2\big)\}_{k\in\bn}$ is increasing. Hence it converges to the fixed point $\mathbf{z}_0(\beta_2)$ of $F(\,\cdot\,;\beta_2)$, that must verify $\mathbf{z}_0(\beta_1)\preceq\mathbf{z}_0(\beta_2)$. Accordingly $\beta_1\leqslant\beta_2$ implies $\mathbf{z}_0(\beta_1)\preceq\mathbf{z}_0(\beta_2)$. Therefore each entry of the vector $\mathbf{z}_0(\beta)$ is an increasing function of $\beta$ and, consequently, $\beta\mapsto\|\mathbf{z}_0(\beta)\|_2$ is increasing. This proves the last assertion in $(c)$ and concludes the proof of the result.
\end{prooftext}

\section{Discussion}
In this paper we have presented a bifurcation analysis for the family of epidemic models known as Microscopic Markov Chain Approach. We prove that the second-order phase transition towards the endemic phase is well captured by a transcritical transition of the dynamical system. Exploiting the analysis of this transition we show that the endemic state is stable and globally attractring for all values of the parameters beyond the critical transition. This result is essential to ground mathematically the numerical scenarios found by finite iterations of the model, and paves the way for further analysis of extensions of the presented model.

\section{Acknowledgments}
A.A., A.G., S.G.\ and J.V.\ acknowledge support from Generalitat de Catalunya (2020PANDE00098). A.A.\ and S.G.\ also acknowledge support from Spanish Ministerio de Ciencia e Innovacion (PID2021-128005NB-C21), Generalitat de Catalunya (2017SGR-896 and PDAD14/20/00001) and Universitat Rovira i Virgili (2019PFR-URV-B2-41). A.G and J.V. also acknowledge support from the Ministry of Science, Innovation and Universities of Spain through the grant MTM2017-86795-C3-2-P. A.A.\ also acknowledges support from ICREA Academia, and the James S. McDonnell Foundation (220020325).

\end{document}